\documentclass[twocolumn,showpacs,,pre,floats,aps,amsmath,amssymb]{revtex4-1}
\usepackage{graphicx}
\usepackage{subfigure}
\usepackage{hyperref} 
\usepackage{bm}
\usepackage{float}
\usepackage{yfonts}

\begin{document}

\title{Dressed tunneling approximation for electronic transport through molecular transistors}
\author{R. Seoane Souto}
\email{ruben.seoane@uam.es}
\author{A. Levy Yeyati}
\author{A. Mart\'{\i}n-Rodero}
\author{R. C. Monreal}
\affiliation{Departamento de F\'{i}sica Te\'{o}rica de la Materia Condensada,\\
Condensed Matter Physics Center (IFIMAC)
and Instituto Nicol\'{a}s Cabrera,
Universidad Aut\'{o}noma de Madrid E-28049 Madrid, Spain}
\date{\today}

\begin{abstract}
A theoretical approach for the non-equilibrium transport properties of 
nanoscale systems coupled to metallic electrodes with strong electron-phonon 
interactions is presented.
It consists in a resummation of the dominant Feynman diagrams 
from the perturbative expansion in the coupling to the leads.
We show that this scheme eliminates the main pathologies found in 
previous simple analytical approaches for the polaronic regime.
The results for the spectral and transport properties are compared
with those from several other approaches for
a wide range of parameters. The method can be formulated in a simple
way to obtain the full counting statistics. Results for the shot 
and thermal noise are presented. 
\end{abstract}

\maketitle

\section{Introduction}
\label{sec:intro}
The effect of localized vibrations (phonons) in the electronic transport 
properties of nanoscale devices is attracting increasing attention 
(for a review see \cite{review}).
Such effects have been identified in different systems like atomic contacts and atomic chains \cite{agrait}, semiconducting
quantum dots \cite{weig}, carbon nanotubes \cite{CNT} and other molecular 
junctions \cite{park,smit,zhitenev}. Other systems in which a strong
electron-phonon coupling can lead to polaronic effects dominating the
electronic transport are organic semiconductors \cite{ortmann,ortmann2}.

In spite of this variety, from the theoretical point of view all these
situations can be qualitatively described by the rather simple 
Anderson-Holstein
model. This model consists of a single resonant electronic level coupled
to fermionic leads and to a localized phonon mode \cite{holstein}. 
Even in the more simple
spinless case, this model corresponds to a non-trivial strongly correlated
system in an out of equilibrium situation. This model can be regarded as
``paradigmatic" of an electronic system interacting with bosonic excitations.
For instance, the same model was proposed by Langreth
to describe the problem of photoemission through core-holes in metals
\cite{langreth}. 

This model has been extensively
analyzed by different theoretical approaches 
\cite{glazman,wingreen,flensberg,mitra} but there is still no
exact solution available except for some limiting cases.  
Comparison with numerically exact methods like numerical renormalization
group or quantum Monte Carlo is possible only for certain
range of parameters both for equilibrium 
\cite{hewson,jeon,cornaglia,liliana} and more recently for 
nonequilibrium situations \cite{rabani,ferdinand1,anders,iterativepi}. 

Within this model one can distinguish between two different regimes 
depending on the strength of the electron-phonon coupling. For 
sufficiently weak coupling a lowest order perturbation theory 
is applicable \cite{viljas,egger,entin}. This
situation is suitable to describe the case of atomic contacts and atomic 
chains \cite{frederiksen,laura}. As the electron-phonon coupling increases
higher order diagrams, including vertex corrections, become of importance
as discussed in Ref. \cite{ness2}.
In the opposite regime, the 
so-called polaronic regime, perturbation
theory breaks down and other type of approaches are necessary
\cite{galperin,braig,vonoppen,alvaro,dong}.

The analysis of the transport properties of this model has been more recently 
extended to the case of noise, 
and, more generally, to its full counting statistics (FCS). 
These studies have been mainly restricted to the
perturbative regime \cite{remi,schmidt,haupt,urban}. 
Although there exist some studies of the noise 
properties in the polaronic regime \cite{galperin2,paperPTA} it is desirable
to develop simple methods to analyze the crossover from the perturbative
to the polaronic case.

Two simple approximations have been proposed to describe the
polaronic regime: the so-called single particle approximation (SPA) and
the polaron tunneling approximation (PTA). Both approaches correspond to 
simple decoupling schemes which allow an analytical evaluation of the 
electronic Green functions. This simplicity has allowed for instance
to extend PTA to analyze the transient behavior of this model 
yielding results in a remarkable good agreement with numerically exact 
ones \cite{Ferdinand}. In spite of their several advantages
both approximations exhibit some pathological features. This is particularly
noticeable in their spectral properties at low frequencies (SPA) and
high frequencies (PTA). Although there exist other methods to describe
this polaronic regime based either on the equation of motion technique
\cite{galperin,carmina1} or other diagrammatic techniques
\cite{carmina2,zazunov,konig,aligia}, 
these methods require a more involved numerical
evaluation. These methods are therefore not easy to extend to more complex
situations like the calculation of the FCS or the analysis of the transient
behavior in the non-stationary case.

The aim of the present work is to develop a simple method
for describing the crossover region from the polaronic to the perturbative
regime. Ideally this method should recover the good features of
SPA and PTA commented above while eliminating their pathologies.
By analyzing the exact perturbation series with respect to the tunneling
to the leads we identify a family of diagrams which gives the dominant 
contribution in the polaronic regime and which can be summed up 
exactly. We will denote this approach as dressed tunneling approximation
(DTA) \cite{master-thesis} as it corresponds to {\it dressing} the leads self-energy with the
polaronic cloud. In spite of being derived for describing the polaronic regime
we show that this approximation gives a reasonable
description of the crossover region while exhibiting an increasing 
deviation from the perturbative results in the corresponding limit.

The manuscript is organized as follows: in Sect. \ref{sec:theory} we
introduce the model Hamiltonian and the basic Green functions formalism
which allows to calculate the different electronic and transport properties.
In Sect. \ref{sec:pert} we analyze the diagrammatic expansion of the
relevant Green functions in the polaronic limit and briefly introduce
the known simple approximations like PTA and SPA. Sect. \ref{sec:DTA}
is devoted to introduce the DTA discussing the arguments for its derivation
and giving the main expressions for the system Green functions. The corresponding
results are described in Sect. \ref{sec:res} where the spectral densities
are compared with other approaches. In this section we also analyze the
DTA results for the transport properties like the current, 
the differential conductance and the noise. Finally, 
we summarize the main results of this work in Sect. \ref{sec:conclusion}.

\section{Model and basic theoretical formulation}
\label{sec:theory}
We consider the simplest spinless Anderson-Holstein 
model in which a single electronic level is coupled to a localized 
vibrational mode. 
Electrons can tunnel from this resonant level into a left (L)
and a right electrode (R). We shall generically refer to this central region, 
which can represent either a molecule, and atomic chain or quantum dot, as
the ``dot" region.
The corresponding Hamiltonian is given by
$H=H_{leads}+H_{dot}+H_T$, with (in natural units, $\hbar=k_B=e=m_e=1$)
\begin{equation}
 H_{dot}=\left[\epsilon_0+\lambda\left(a^\dagger+a\right)\right]d^\dagger d+\omega_0\;a^\dagger a \;,
  \label{Hv}
\end{equation}
where $\epsilon_0$ is the bare electronic level, $\lambda$ is the
electron-phonon coupling constant and $\omega_0$ is the frequency of the 
localized vibration. The electron (phonon) creation operator in the dot
is denoted by
$d^{\dagger}$ ($a^{\dagger}$). On the other hand,
$H_{leads}=\sum_{j k}\epsilon_{j k} c_{j k}^\dagger c_{j k}$
corresponds to the non-interacting leads Hamiltonian ($j\equiv L,R$) where 
$\epsilon_{j k}$ are the leads electron energies and 
$c^{\dagger}_{j k}$ are the corresponding creation 
operators. The bias voltage applied to the junction is imposed by shifting the 
chemical potential of the electrodes $V=\mu_L-\mu_R$.\newline
\hspace*{5mm}The tunneling processes are described by
\begin{equation}
  H_T=\sum_{k} \left(t_{L k} \;c_{L k}^{\dagger}\;d+t_{R k} \;c_{R k}^{\dagger}\;d+\mbox{h.c.}\right) \; ,
\end{equation}
where $t_{j k}$ are the tunneling amplitudes. 

To address the polaronic regime it is convenient to perform the so-called
Lang-Firsov unitary transformation \cite{LangFirsov} which allows to eliminate
the linear term in the electron-phonon coupling \cite{Mahan}
\begin{equation}
 \tilde{H}=S H S^\dagger , \quad S=e^{g d^\dagger d (a^\dagger - a)} ,\quad g=\frac{\lambda}{\omega_0} \;.
\end{equation}

Using this transformation
\begin{equation}
 \tilde{H}_{dot} =\tilde{\epsilon}\;d^\dagger\;d \; + \; \omega_0 a^\dagger a 
\;,
\end{equation}
where $\tilde{\epsilon}=\epsilon_0-\lambda^2/\omega_0$. The tunneling 
Hamiltonian is transformed as
\begin{equation}
 \tilde{H}_T=\sum_{k} \left(t_{L k}\;c_{L k}^{\dagger}\;X d+t_{R k} \;c_{R k}^{\dagger}\;X d+\mbox{h.c.}\right) \; .
\end{equation}
where $X = \exp{\left[g (a - a^{\dagger})\right]}$ is the phonon cloud 
operator. On the other hand, the free leads Hamiltonian remains invariant.
For later use it is useful to introduce the tunneling rates $\Gamma_{j} 
= \mbox{Im} \sum_k |t_{j k}|^2/(\omega - i0^+ - \epsilon_{j k})$
which are approximated by constants in the so-called wide band approximation.

To deal with the transport properties of this model it is convenient to use
the Keldysh nonequilibrium formalism \cite{Keldysh}. The basic quantity 
required to calculate the electronic and transport properties are the
dot Green functions
\begin{equation}
 G^{\alpha\beta}(t,t')=-i\left\langle T_{\cal{C}}\{X(t) d(t) X^{\dagger}(t')
d^\dagger(t')\}\right\rangle\;, 
  \label{G0}
\end{equation}
where $T_{\cal{C}}$ is the time ordering operator in the Keldsyh contour 
and $\alpha, \beta  \equiv +,-$ denote the different branches of the
contour. 

A slight modification in the Keldysh formulation allows to address directly
the noise properties of the system and more generally its FCS \cite{FCS,FCS-2}. 
This is achieved by introducing a ``counting-field" $\nu$, which changes
sign on the two branches of the contour and which enters as
a phase factor modulating the tunnel Hamiltonian. 
As the current is conserved 
in our two terminal device one can choose to introduce the counting field
in either the left or the right tunneling term. For definiteness we 
choose to include it in the left and accordingly we define
\begin{equation}
\tilde{H}^{\nu}_T =\sum_{k} \left(e^{i\nu/2} t_{L k}\;c_{L k}^{\dagger}\;X d+t_{R k} \;c_{R k}^{\dagger}\;X d+\mbox{h.c.}\right) \;,
\end{equation}
and the Keldysh Green functions (GFs) in the presence of the counting field are
\begin{widetext}
\begin{equation}
 G^{\alpha\beta(\nu)}(t,t')=-i\left\langle T_{\cal{C}} \{X(t) d(t) X^{\dagger}(t')
d^\dagger(t') e^{-i\int_{\cal{C}} \tilde{H}^{\nu}_T (\tau) d\tau} \}
\right\rangle_0\;, 
  \label{G0nu}
\end{equation}
\end{widetext}
where the subscript $0$ indicates averaging over the 
states of the uncoupled Hamiltonians $\tilde{H}_{dot}$ and $H_{leads}$.
More generally, the FCS can be obtained from a Cumulant
Generating Function (CGF), 
\begin{equation}
 \chi(\nu)=\left\langle T_{\cal{C}} \;e^{-i\int_c dt \tilde{H}^{\nu}_T (t)}\right\rangle_0 \; .
\end{equation}

Formally $\chi(\nu)$ can be expanded as $\chi(\nu)=\sum_q e^{iq\nu}P_q$, where 
$P_q$ is the probability of a charge $q$ being transferred through the system. 
So, the cumulants can be computed using
\begin{equation}
 \left\langle\delta^n \;q\right\rangle=\left.(-i)^n\;\frac{\partial^n}{\partial\;\nu^n}\ln\chi(\nu)\right|_{\nu=0}.
  \label{cumulants}
\end{equation}

In a stationary situation the first cumulant ($n=1$) in (\ref{cumulants}) corresponds to the mean current
\begin{eqnarray}
I_L &=& \int \frac{d\omega}{2\pi} \sum_k \left[ 
t_{Lk} g^{+-}_{L k}(\omega)G^{-+}(\omega)  
\right. \nonumber\\
&& - \left. t^*_{Lk} g^{-+}_{L k}(\omega)G^{+-}(\omega) \right] ,
\label{generalI}
\end{eqnarray}
where $g^{\alpha\beta}_{jĸ}$ are the isolated leads GFs.
A symmetrized expression of the current involving only the dot spectral density
can be deduced using current conservation, leading to \cite{Meir}
\begin{equation}
 I=\frac{8\Gamma_L\Gamma_R}{\Gamma} \int d\omega \left[f_L(\omega)-f_R(\omega)\right] A(\omega)\; ,
 \label{ispectral}
\end{equation}
where $\Gamma=\Gamma_L+\Gamma_R$ and the spectral function
$A(\omega)=-\mbox{Im}(G^{R}(\omega))/\pi$, $G^R$ being the retarded
dot Green function

On the other hand the second cumulant corresponds to the current noise 
and can be written as
\begin{eqnarray}
S_L &=& i\int\frac{d\omega}{2\pi} \sum_k 
\frac{\partial}{\partial\nu}\left[t_{L k} G^{-+(\nu)}(\omega) 
g_{L k}^{+-}(\omega)\;e^{i\nu} \right. \nonumber\\
&& \left. \left.-t^*_{L k} G^{+-(\nu)}(\omega) g_{L k}^{-+}(\omega)\;e^{-i\nu}\right]\right|_{\nu=0} .
  \label{generalnoise}
\end{eqnarray}

Symmetrized expressions for the noise in different approximations 
are given in Appendix \ref{appendix}.

It is also convenient to define the unperturbed polaron correlator in Keldysh space
\begin{equation}
 \Lambda^{\alpha\beta}(t,t')=\left\langle T_C e^{g\left(a(t)-a^\dagger(t)\right)}e^{-g\left(a(t')-a^\dagger(t')\right)}\right\rangle_0 .
\end{equation}

\section{Diagrammatic expansions in the polaronic regime}
\label{sec:pert}

\begin{figure}
\begin{center}
    \begin{minipage}{1\linewidth}
      \includegraphics[width=1\textwidth]{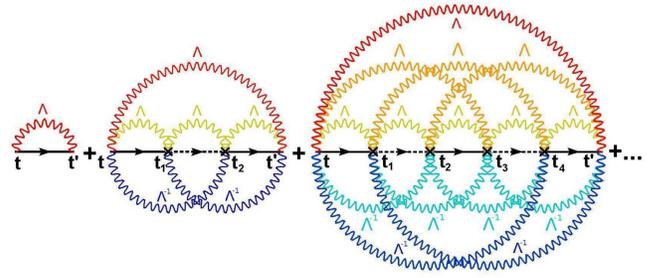}
    \end{minipage}
\end{center}
\caption{(Color online) Lowest order Feynman diagrams of the perturbative 
expansion in the tunneling Hamiltonian. The solid lines represents the free dot GF ($G_0$), 
dashed lines represent the free lead GF ($g_k$, $k=L,R$), the wavy lines 
the polaron correlator ($\Lambda$) and the crosses the hopping events.}
\label{total_pert}
\end{figure}

An exact solution to the problem of determining the GFs entering in the
calculations of the various transport properties is still unknown.
In the polaronic regime a perturbative expansion in the hopping to the
leads would be appropriate.
The lowest order diagrams in this expansion are shown in Fig. 
\ref{total_pert}.

Therefore, the natural starting point 
for studying the system in this regime is provided by the so-called
atomic limit \cite{more-atomic}. This is defined as the limit when the tunneling rates between
the dot and the
leads tend to zero. The Green functions in this limit can be calculated
exactly and correspond to  the zero order term in the expansion depicted in
Fig. \ref{total_pert}, 
and thus its Keldysh components can be computed as
 ${G^{(0)\alpha\beta}(\omega)=G_{0}^{\alpha\beta}(\omega)\otimes\Lambda^{\alpha\beta}(\omega)}$, where $G_0^{\alpha\beta}$ are the bare
dot GFs and $\otimes$ represents the convolution product.
In frequency domain its retarded component has the form
\begin{equation}
 G^{(0) R}(\omega)=\sum_{k=-\infty}^{\infty}\frac{\alpha_k n_0 +\alpha_{-k} (1-n_0)}{\omega-\tilde{\epsilon}+k\;\omega_0+i\eta} \; ,
  \label{zerorder}
\end{equation}
where $n_0$ represents the average occupation number of the level of the dot, 
$\eta$ is an infinitesimal and $\alpha_k$ is a coefficient that, at finite 
temperature can be written as
\begin{equation}
 \alpha_k=e^{-g^2(2\;n_p+1)}\;I_k\left(2g^2\sqrt{n_p(1+n_p)}\right)\;e^{k\beta\omega_0/2} ,
\label{alpha}
\end{equation}
$I_k$ being the modified Bessel function of the first kind which is symmetric 
in the $k$ argument ($I_k=I_{-k}$) and $n_p$ is the Bose factor
$1/(e^{\beta\omega_0} - 1)$ with $\beta = 1/T$. At zero temperature this coefficient 
can be simplified as
\begin{equation}
 \alpha_k =\left\{\begin{array}{rrr}
e^{-g^2}\frac{g^{2k}}{k!}&\mbox{if }k\geq0\\
0&\mbox{if }k<0
\end{array}\right. 
\label{alpha2}
\end{equation}

There exist in the literature two simple ways to include the effects of
finite tunneling to the leads starting from the atomic limit. These are 
the so called Polaron Tunneling Approximation (PTA) and the Single Particle 
Approximation (SPA), which are briefly described below.

\begin{figure}
    \begin{minipage}{1\linewidth}
      \includegraphics[width=1\textwidth]{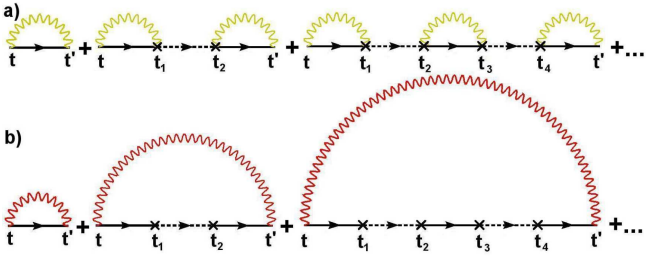}
    \end{minipage}
  \caption{(Color online) Feynman diagrams for (a) PTA (b) DSPA. 
In both cases, the zero order corresponds to the atomic limit GF
(\ref{zerorder}).}
\label{PTA}
\end{figure}

Within PTA the phonons are assumed to be excited and deexcited 
instantaneously when the electrons tunnel from the leads to the dot \cite{paperPTA}. 
Diagrammatically, this approximation corresponds to summing up the
series depicted in Fig. \ref{PTA} (a), i.e. can be expressed in a Dyson-like  
equation in Keldysh space
(for simplicity we concentrate in the $\nu=0$ case for the 
discussion within this and the next section)
\begin{equation}
\textbf{G}_{PTA}=\textbf{G}^{(0)} +
\textbf{G}^{(0)} {\bf \Sigma}_0 \textbf{G}_{PTA},
\label{Dyson}
\end{equation}
where the self-energy ${\bf \Sigma}_0={\bf \Sigma}_{0L}+{\bf \Sigma}_{0R}$, 
with
\begin{equation}
{\bf \Sigma}_{0j}(\omega) = i \Gamma_j \left(\begin{array}{cc}  
2f_j(\omega)-1 &  -2 f_j(\omega) \nonumber\\
-2 (f_j(\omega) - 1) &  2f_j(\omega)-1 
\end{array} \right).
\end{equation}

This approximation provides a good description of the spectral properties
at low energies and for situations close to half-filling
\cite{Ferdinand}. In particular it satisfies the
Friedel sum rule (FSR) connecting the spectral density at zero energy 
with the dot charge, which implies that for the symmetric case
($\tilde{\epsilon} = 0$) $A_{PTA}(0) = 1/\pi\Gamma$ \cite{alvaro}.
However, the spectral density at higher energies
is somewhat pathological as it exhibits phonon side band peaks of vanishing 
width but with a constant height. On the other hand, as has been shown in
Ref. \cite{Ferdinand} this approximation provides a good description of the
non-stationary evolution of the model at short time scales. 

The SPA provides another simple picture of the polaronic
regime. The general idea of this approximation is to directly decouple the
electronic and the polaron degrees of freedom in the dot GFs. 
In the simplest form of this approximation \cite{flensberg} the retarded GF 
is given by
\begin{equation}
G^{R}_{SPA}(\omega) = \sum_{k=-\infty}^{\infty} \frac{\alpha_k n_0 + \alpha_{-k}
(1 - n_0)}{\omega - \tilde{\epsilon} + k \omega_0 + i \Gamma} \;.
\label{SPA}
\end{equation}

This expression is formally equivalent to broaden the poles of the atomic
GF of Eq. (\ref{zerorder}) with the bare (frequency independent) 
tunneling rates to the leads. From this expression it is clear that this
approximation does not satisfy the expected behavior at low frequencies as
it does not fulfill the FSR nor reproduce the polaronic narrowing of the
resonances close to the Fermi level. However, it does not exhibit the 
pathological behavior of the side band peaks at higher energies characteristic
of PTA and recovers the exact results in the limit of a fully occupied or 
fully empty dot \cite{SPA-exact,glazman}.

In a more transparent diagrammatic way the SPA decoupling scheme
corresponds to the Feynman diagrams showed in 
Fig. \ref{PTA} (b) in which the bare dot Green function is dressed
with the tunneling self-energy up
to infinite order and then convoluted with the polaron correlator $\Lambda$
\cite{zazunov}. The resulting GFs are, however, not completely equivalent to
the ansatz of Eq. (\ref{SPA}). The Keldysh components within this
diagrammatic SPA (DSPA) are given by
\begin{eqnarray}
  G_{DSPA}^{+-}(\omega)=-\sum^{\infty}_{k=-\infty}\alpha_k \frac{\Sigma^{+-}_0(\omega+k\;\omega_0)}{\textfrak{D}(\omega+k\;\omega_0)}\nonumber\\
  G_{DSPA}^{-+}(\omega)=-\sum^{\infty}_{k=-\infty}\alpha_{-k} \frac{\Sigma^{-+}_0(\omega+k\;\omega_0)}{\textfrak{D}(\omega+k\;\omega_0)} \; ,
\end{eqnarray}
where $\textfrak{D}(\omega) =(\omega-\tilde{\epsilon})^2 + \Gamma^2$.

\section{Dressed Tunneling Approximation}
\label{sec:DTA}
\begin{figure}
    \begin{minipage}{1\linewidth}
      \includegraphics[width=1\textwidth]{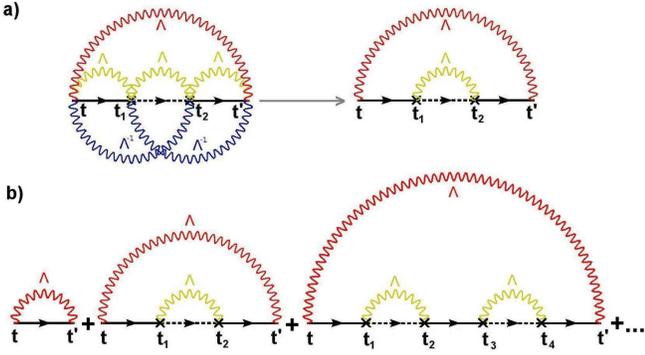}
    \end{minipage}
  \caption{(Color online) Feynman diagrams for the DTA approximation. Panel a) 
indicates the simplifying approximation on the second order diagram and
panel b) represents the diagrammatic series which is included within DTA.
}
  \label{MSPA}
\end{figure}

It is thus desirable to develop a simple approximation that would exhibit the
properties of the PTA at low energies and of SPA at high energies.
For this purpose let us analyze the full 
diagrammatic expansion illustrated in Fig. \ref{total_pert}, taking
as an example the second order diagram represented again in Fig. 
\ref{MSPA} (a).  In the evaluation of this diagram there appear
products of polaron correlators
of the type $\Lambda(t,t_1) \Lambda^{-1}(t,t_2)$,
where the time arguments $t_1$ and $t_2$ correspond to the exit and entrance
of the electrons from the dot to the leads.
As in the limit of strong electron-phonon coupling the lifetime of the electronic states 
in the dot is much larger than the one in the electrodes,
it is then reasonable to make the approximation 
$\Lambda(t,t_1) \Lambda^{-1}(t,t_2) \sim 1$ (see Fig.\ref{MSPA} (a)). 
This can be more rigorously justified from the fact that in the 
wide band approximation the retarded leads self-energies are localized in
time representation, i.e. 
$\Sigma^R_{0j}(t,t') \propto \theta(t-t')\delta(t-t')$.
With this prescription the diagrammatic expansion reduces to the one 
illustrated in Fig. \ref{MSPA} (b) which can be evaluated exactly. 
One should notice that the cancellation of the ``crossing" polaron lines
in the diagrammatic expansion implies that vertex corrections can be
neglected in this limit. An approximate evaluation of these corrections
in the $\lambda/\omega_0 <1$ regime
was undertaken in Ref. \cite{zazunov}.

As can be observed, the resulting approximation is equivalent to 
{\it dressing} the leads self-energies within DSPA with the polaron 
correlators, i.e. ${\tilde{\Sigma}^{\alpha\beta}(\omega)=
\Sigma^{\alpha\beta}_0(\omega)\otimes\Lambda^{\alpha\beta}(\omega)}$. 
In this way, the self-energy components can be written as
\begin{eqnarray}
\tilde{\Sigma}^{+-}_{L,R}(\omega)=\sum^{\infty}_{k=-\infty}\alpha_k 
\Sigma^{+-}_{0\;L,R}(\omega+k\;\omega_0)\nonumber\\
\tilde{\Sigma}^{-+}_{L,R}(\omega)=\sum^{\infty}_{k=-\infty}\alpha_{-k} 
\Sigma^{-+}_{0\;L,R}(\omega+k\;\omega_0) \;.
\end{eqnarray}

Within the wide-band approximation these
self-energy components are purely imaginary quantities.
The resulting GFs can then be straight-forwardly evaluated as

\begin{eqnarray}
  G_{DTA}^{+-}(\omega)=-\sum^{\infty}_{k=-\infty}\alpha_k \frac{\tilde{\Sigma}^{+-}(\omega+k\;\omega_0)}{\tilde{\textfrak{D}}(\omega+k\;\omega_0)}\nonumber\\
  G_{DTA}^{-+}(\omega)=-\sum^{\infty}_{k=-\infty}\alpha_{-k} \frac{\tilde{\Sigma}^{-+}(\omega+k\;\omega_0)}{\tilde{\textfrak{D}}(\omega+k\;\omega_0)}
  \label{GF_DTA}
\end{eqnarray}
where  $\tilde{\Sigma}^{\alpha\beta} = \tilde{\Sigma}^{\alpha\beta}_L +
\tilde{\Sigma}^{\alpha\beta}_R$, and
\begin{eqnarray}
 \tilde{\textfrak{D}}(\omega)=\left|\omega-\tilde{\epsilon}-\tilde{\Sigma}^{R}(\omega)\right|^2 ,
\end{eqnarray}
where 
\begin{eqnarray}
 \tilde{\Sigma}^R(\omega)=\sum^{\infty}_{\substack{k=-\infty\\ j=L,R}}\frac{i\;\alpha_k}{2\pi}\int d\omega'\left[\frac{\Sigma_{0j}^{+-}(\omega')}{\omega+k\;\omega_0-\omega'+i\eta}\right.\nonumber\\
+ \left.\frac{\Sigma_{0j}^{-+}(\omega')}{\omega-k\;\omega_0-\omega'+i\eta}\right].
\end{eqnarray}

With these components, the spectral function can be determined as
\begin{equation}
 A_{DTA}(\omega) =\frac{1}{2\pi i} \left[G^{+-}_{DTA}(\omega) - 
G^{-+}_{DTA}(\omega)\right] \;.
\end{equation}

It should be noticed that a similar approach but derived from a
decoupling procedure within the equation of motion of the system 
GFs was presented recently in Ref. \cite{dong}. We also point out that
in all the preceding approximations the basic assumption of having a
equilibrium phonon distribution was made for the evaluation of the
polaron correlators.

A simpler version of this approximation can be obtained within the
same spirit as in the SPA discussed in the previous section. 
Within this approximation
(that we call approximated DTA (ADTA)), $G^R$ can be written as
\begin{equation}
 G^R_{ADTA}(\omega) =
\sum_{k=-\infty}^{\infty}\frac{\alpha_k\;n_0+\alpha_{-k}\;(1-n_0)}
{\omega+k\omega_0-\tilde{\epsilon}-\tilde{\Sigma}^R(\omega+k\omega_0)} \;.
\end{equation}

From this expression it is clear that within this approximation the 
pole structure of the atomic limit is preserved but with a broadening 
determined by $\mbox{Im}\tilde{\Sigma}^R$. As for large frequencies this
effective broadening tends to $\Gamma$ one recovers the SPA result in this 
limit. 
However, this effective broadening is strongly reduced with respect
to $\Gamma$ for low energies. In fact, for $\omega \rightarrow 0$ and
for the symmetric case $|\mbox{Im}\tilde{\Sigma}^R| \rightarrow e^{-g^2} \Gamma$,
thus yielding the correct polaronic reduction in the resonance width at
the Fermi level \cite{carmina1}. 

\section{Results}
\label{sec:res}

This section contains the predictions of the DTA for different physical
quantities compared with other approaches. 
 
\subsection{Spectral density}

\label{subsec::spectral}
\begin{figure}
     \begin{minipage}{1\linewidth}
      \includegraphics[width=1.0\textwidth]{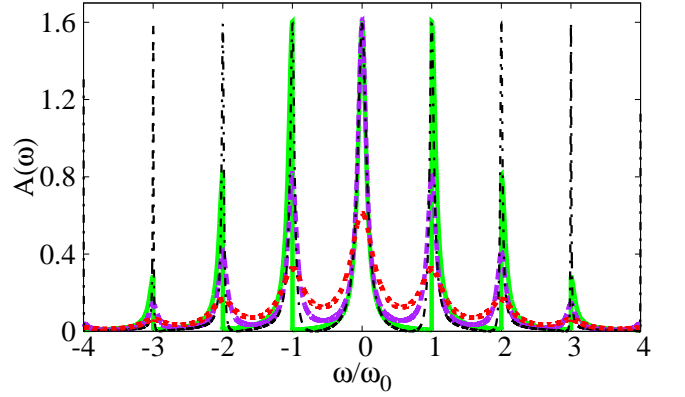}
     \end{minipage}
  \caption{(Color online) Spectral density for 
the symmetric case, zero bias voltage and zero temperature with 
$\Gamma=0.2\;\omega_0$ and $g=1.0$. The results correspond to the 
different approximations: DTA (full line, green or light gray), 
ADTA (full line, violet or dark gray), SPA (dotted line) and PTA (dashed line).}
\label{DTA-vs-SPA-PTA}
\end{figure}

We first analyze the results for the dot spectral density $A(\omega)$. 
Fig. \ref{DTA-vs-SPA-PTA} shows the comparison of the DTA results
for $A(\omega)$ with those from PTA and SPA for an electron-hole
and left-right symmetric case in the polaronic regime. 
As can be observed, while the DTA and the PTA results
tend to coincide at low energies (central resonance), they increasingly 
deviate at higher order resonances. In contrast, for these higher order 
resonances the DTA spectral density gradually converges to the SPA one. 
Therefore, as commented above, DTA contains the good features of both
approximations but without their pathologies. The same conclusion
is valid for the simpler ADTA method, as can be seen in Fig. \ref{DTA-vs-SPA-PTA}. 

\begin{figure}
     \begin{minipage}{1.0\linewidth}
      \includegraphics[width=1.0\textwidth]{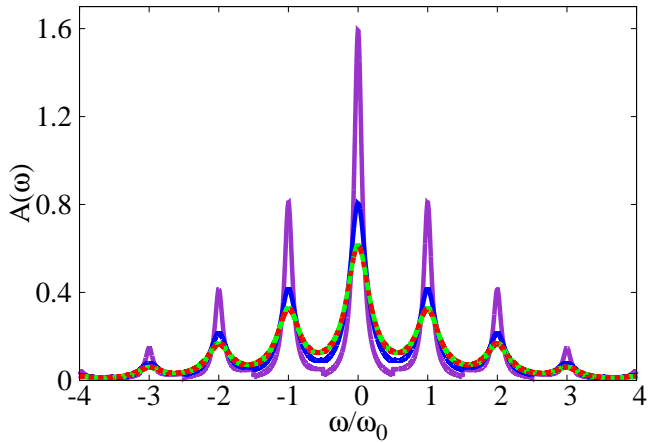}
     \end{minipage}
\caption{(Color online) Evolution of the spectral density at zero temperature with increasing 
voltage within DTA. From top to bottom $V = 1$, 3, 9 $\omega_0$. 
The other parameters
as in lower panel of Fig. \ref{interpolation}. The dotted line
corresponds to the SPA result which is voltage independent.}
\label{spectral-vs-voltage}
\end{figure}

In Fig. \ref{spectral-vs-voltage} we analyze the evolution of the
spectral density with the applied bias voltage. As can be observed, the
main effect of the applied bias is to gradually reduce the height of the
phonon peaks. Remarkably, the DTA spectral density appears to evolve towards
the SPA one which is voltage independent (indicated by the dotted 
curve in Fig. \ref{spectral-vs-voltage}). As large voltages 
correspond to high energies one would expect that SPA
should become exact in the limit $V \rightarrow \infty$. A
similar convergence to the SPA is obtained in the limit $|\tilde{\epsilon}|
\rightarrow \infty$, corresponding to the exact result for
a fully empty or a fully dot case \cite{SPA-exact}.

\begin{figure}
     \begin{minipage}{1.0\linewidth}
      \includegraphics[width=1\textwidth]{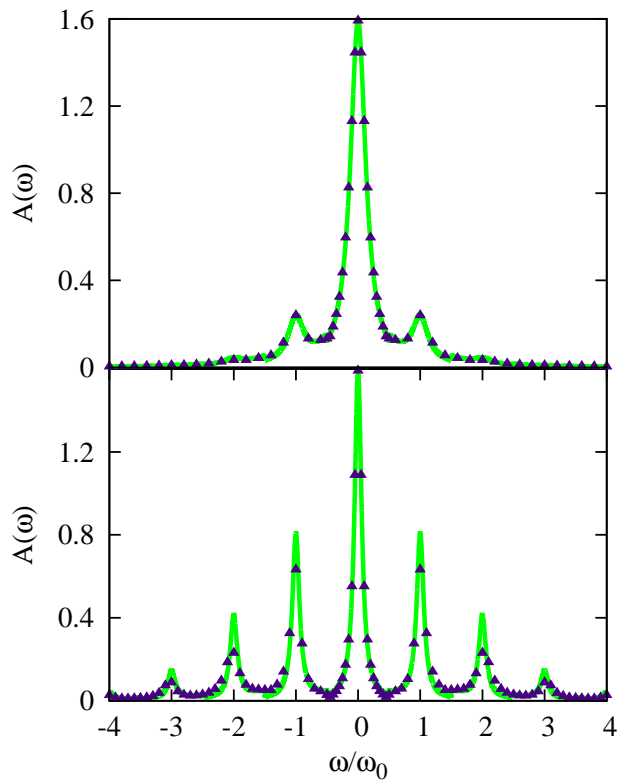}
     \end{minipage}
\caption{(Color online) Spectral density at zero temperature in the DTA (full line) and the 
ISA (triangles) from Ref. \cite{carmina2} for the symmetric case with
$\Gamma=0.2\omega_0$, $V=\omega_0$ and two values of the 
coupling constant $g=0.5$ (upper panel) and $g=1$ (lower panel).}
\label{interpolation}
\end{figure}

Another interesting property of DTA is that it reasonably describes the
transition from the polaronic to the weak electron-phonon coupling regimes.
This is illustrated in Fig. \ref{interpolation} in which the DTA spectral
density is shown for two values of the parameter $g = 1$ (lower panel)
and $g=0.5$ (upper panel). For comparison we also show in these
plots the corresponding results obtained by the interpolative self-energy 
approach (ISA) of Ref. \cite{us1993} 
and extended to the non-equilibrium Holstein model in Ref. \cite{carmina2}, 
which is constructed 
in order to interpolate between the second-order perturbation theory 
and the atomic limit. It is remarkable that the two approximations which
are derived following such different criteria would so closely
coincide in both regimes. 

\subsection{Current and Noise}
\label{subsec::current}

\begin{figure}
    \begin{minipage}{1\linewidth}
      \includegraphics[width=1\textwidth]{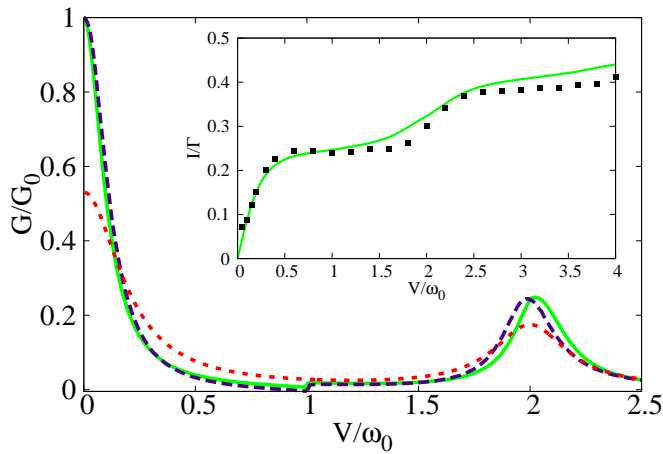}
    \end{minipage}
  \caption{(Color online) Zero temperature differential conductance within 
DTA (full line, green or light gray), 
ISA (dotted line) and SPA (dots) for 
$\tilde{\epsilon}=0$, $\Gamma=0.1 \omega_0$
and $g=0.8$. The inset shows the current for the same parameters 
but with a finite temperature $T=0.04 \omega_0$ for comparison
with diagrammatic Monte Carlo data (full squares) from \cite{rabani}.} 
  \label{intMC}
\end{figure}

We analyze in this subsection the results from the DTA for several
transport properties. 
As shown in the inset of Fig. \ref{intMC} for moderate values of $g$
($g \sim 0.8$) the current-voltage characteristic starts to exhibit a step-like 
behavior.  
For the electron-hole and left-right symmetric case shown in Fig. \ref{intMC}
the most pronounced features appear at $V \sim 2n\omega_0$ \cite{paperPTA,carmina2,anders}.
It is interesting to note that the DTA results for the current
quantitatively agree in this range of parameters with numerically exact
results from diagrammatic MC calculations from Ref. \cite{rabani}, 
indicated by the symbols in the inset of Fig. \ref{intMC}. It should
be also mentioned that, as shown in Appendix \ref{appendix}, DTA 
fulfills the current conservation condition. This condition is trivially
fulfilled by PTA where no inelastic processes are included, but not
for instance by DSPA.  In the case of DSPA the violation of current
conservation can be demonstrated explicitly (see Appendix \ref{appendix}).
However, as SPA and ADTA consist in an ansatz for the retarded GFs
the left and right currents cannot be calculated separately but just
using the symmetrized expression of Eq. (\ref{ispectral}).
Their non-conserving character can be inferred nevertheless from the
violation of the FSR.

In order to have a more detailed analysis of the features in the IV 
characteristics it is convenient to calculate the differential conductance. 
This quantity is represented in the main panel of Fig. \ref{intMC} for the 
same parameters as for the inset. We also show for comparison the 
corresponding results for the SPA and the ISA. Several features are worth 
noticing: 1) the zero bias conductance within DTA reaches the unitary limit
as it corresponds to an electron-hole symmetric case. This condition,
which is directly related to the FSR, is also
fulfilled by ISA but not by SPA which yields a smaller conductance value.
2) There appears a conductance step at $V\sim \omega_0$. This step corresponds
to the onset of inelastic processes due to phonon emission, which is absent
within SPA (neither PTA, not shown in Fig. \ref{intMC},
exhibits this feature), and 3) There appears a more pronounced feature at
$V \sim 2\omega_0$ corresponding to the side-band peaks in the spectral
density. It should be noticed that the precise shape of this feature
is extremely sensitive to the presence of a finite broadening of the 
logarithmic singularities in the real part of the electron self-energies. In fact,
a finite broadening leads to a dip in the differential conductance 
at $V = 2\omega_0$ \cite{dong} which however tends to disappear as
the broadening is reduced to zero. 

\begin{figure}
    \begin{minipage}{1\linewidth}
      \includegraphics[width=1\textwidth]{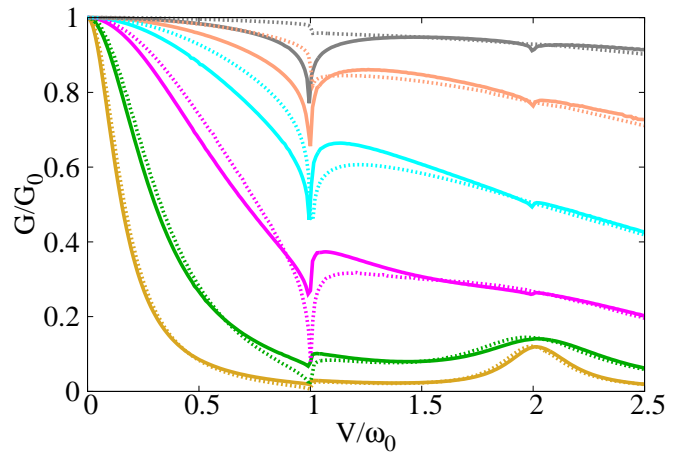}
    \end{minipage}
\caption{(Color online) Zero temperature conductance within DTA for 
$\tilde{\epsilon}=0$, $g=0.5$ and (from bottom to top) 
$\Gamma=0.1$, 0.2, 0.5, 1.0, 2.0 and 4.0 $\omega_0$. The dashed lines
show the corresponding results for ISA.}
  \label{increase-gamma}
\end{figure}

The inelastic features at $V \sim \omega_0$ become more pronounced as $\Gamma$
is increased. This is illustrated in Fig. \ref{increase-gamma} where the
conductance is shown for fixed $g$ and increasing values of $\Gamma$. 
An interesting issue, which has been addressed repeatedly in the literature
is the transition from a step up to a step down in the conductance at the
inelastic threshold. For instance, this transition was analyzed in the
perturbative regime for the electron-phonon coupling in Refs. 
\cite{laura,frederiksen}. 
As shown in Fig. \ref{increase-gamma} the DTA 
fairly reproduces the step up feature at small values of $\Gamma$ but
the transition to the step down behavior at large $\Gamma$ is somewhat
masked by the presence of the logarithmic singularity in the real part of
the electron self-energy. For comparison we show in Fig. \ref{increase-gamma}
the corresponding results obtained with ISA which by construction
reproduces the expected step down feature in the large $\Gamma$ in 
agreement with perturbation theory in $g$. 

\begin{figure}
    \begin{minipage}{1\linewidth}
      \includegraphics[width=1\textwidth]{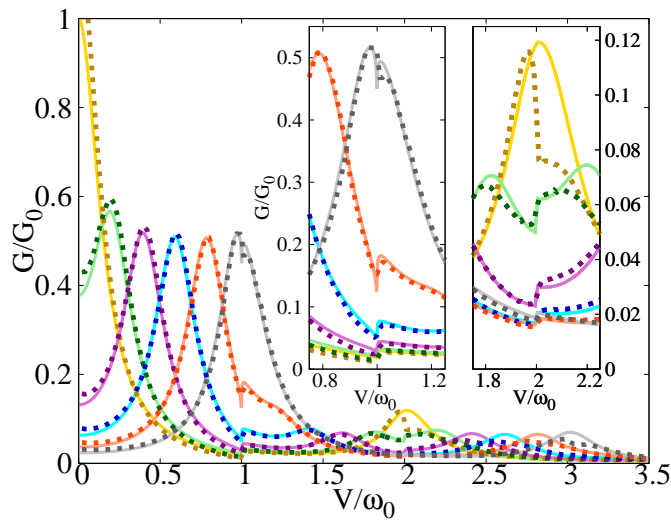}
    \end{minipage}
\caption{(Color online) Zero temperature conductance within 
DTA for $g=1$, $\Gamma = 0.1 \omega_0$ and (from
left to right) increasing values of $\tilde{\epsilon}= 0.0$,
0.1, 0.2, 0.3, 0.4 and $0.5 \omega_0$. The dashed lines
show the corresponding results for the EOM method of 
Ref. \cite{carmina1}. The insets correspond 
to a blow up of the first and second phonon resonance.}
  \label{DTA-EOM}
\end{figure}

The behavior of the differential conductance as $\tilde{\epsilon}$ is
varied is shown in Fig. \ref{DTA-EOM}. It is interesting to analyze the
evolution of the features at the inelastic threshold $V = \omega_0$.
As can be observed in the first inset of Fig. \ref{DTA-EOM},
the initial step up feature for the symmetric case
evolves into a step down as $\tilde{\epsilon}$ approaches $\omega_0/2$ where
the elastic resonance condition $V/2 = \tilde{\epsilon}$
coincide with the inelastic threshold. In Fig. \ref{DTA-EOM} we compare
the DTA results with those of the EOM method obtained in Ref. \cite{carmina1}.
As can be observed there is a remarkable agreement between the two
methods in this range of parameters. Additionally, the results exhibit
a second inelastic threshold at $V \sim 2\omega_0$ which can be more
clearly appreciated when $\tilde{\epsilon}$ is shifted from zero energy  
(see right inset of Fig. \ref{DTA-EOM}). This feature cannot be recovered
by other methods like PTA, SPA or ISA.

\begin{figure}
    \begin{minipage}{1\linewidth}
      \includegraphics[width=1\textwidth]{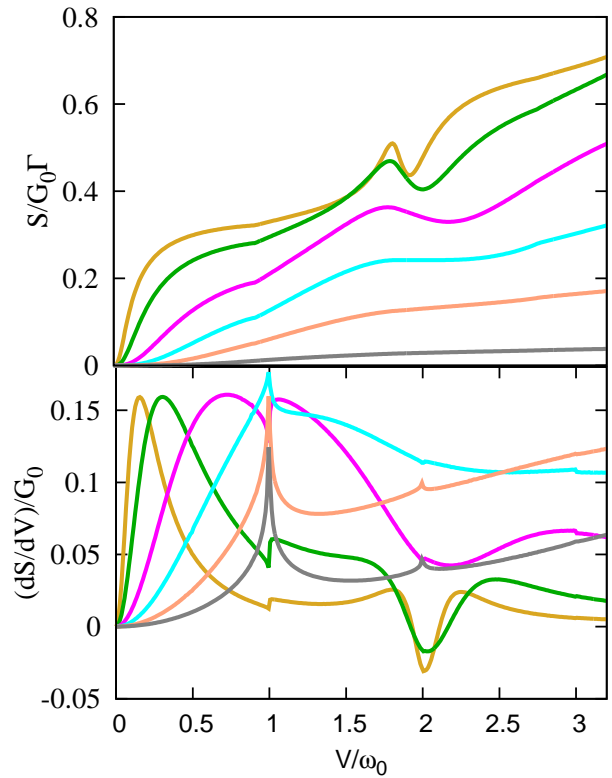}
    \end{minipage}
  \caption{(Color online) Zero frequency noise within DTA for the same parameters
as in Fig. \ref{increase-gamma}. The lower panel corresponds to the differential
noise $\partial S/\partial V$.
From top to bottom in the upper panel
$\Gamma = 0.1$, 0.2, 0.5, 1.0, 2.0 and 4.0 $\omega_0$.}
  \label{noise-vs-V}
\end{figure}

We next analyze the results for the current noise within DTA.
Fig. \ref{noise-vs-V} 
shows the noise and the differential noise, $\partial S/\partial V$ 
as a function of voltage for increasing values of $\Gamma$ at
fixed $g$ for the same
parameter choice as in Fig. \ref{increase-gamma} for the differential
conductance. There is an overall behavior of $\partial S/\partial V$ which 
is maintained for all values of $\Gamma$: it starts from a zero value at
$V=0$ as it corresponds to a perfect transmitting channel at zero temperature,
there is then a maximum at around $V \sim 2\Gamma$ followed by a decay as 
expected for a Lorentzian resonance. In addition the noise exhibits 
features at $V \sim n \omega_0$ as the conductance. However, there 
appear interesting differences. For instance, the feature at the inelastic
threshold at $V \sim \omega_0$ evolves as a function of $\Gamma$
from a step up at small values, 
to a step down at intermediate ones and eventually again to a step
up at large $\Gamma$.  As in the case of the conductance the features
at large $\Gamma$ are masked by the logarithmic singularities in the real part
of the self-energies and can only be identified by analyzing the behavior of
the noise on the neighborhood of the inelastic threshold. 
This double change of sign in the step is qualitatively
in agreement with previous analysis of the noise in the perturbative 
regime \cite{remi,schmidt,haupt}
and which has been partially confirmed by experimental results
for transport through small molecules \cite{kumar}. 

As a final issue we discuss the zero-bias limit for the noise. 
In this limit the noise is purely due to thermal fluctuations and one
should recover the fluctuation-dissipation theorem (FDT), stating that
$S(0) = 4 T G(0)$. This can be proved analytically within DTA (and also
for the DSPA) as shown in Appendix \ref{appendix}.
Fig. \ref{noise-vs-T} illustrates the behavior of the thermal 
noise as a function of temperature for increasing values of $\Gamma$
at fixed $g$. As can be observed, all curves converge to the value 
$S(0)/4TG_0 = 1$ in the zero-temperature limit, thus indicating the
fulfillment of FDT within this approach.

\begin{figure}
    \begin{minipage}{1\linewidth}
      \includegraphics[width=1\textwidth]{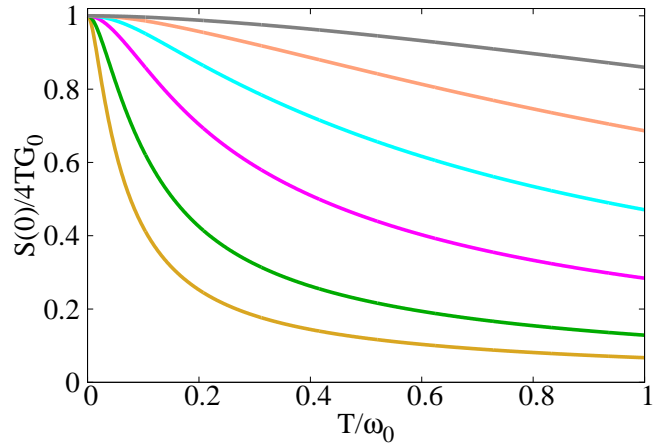}
    \end{minipage}
  \caption{(Color online) Thermal noise as a function of temperature
within DTA for the same parameters as in Figs. \ref{increase-gamma} and
\ref{noise-vs-V}. The noise is normalized as $S(0)/4TG_0$ in such a way
as to illustrate the fulfillment of the FDT in the zero-temperature limit.}
\label{noise-vs-T}
\end{figure}

\section{Conclusion}
\label{sec:conclusion}
In this work we have presented and analyzed a theoretical approach for the non-equilibrium transport properties of nanoscale systems coupled to
metallic electrodes with strong electron-phonon interactions.
We have shown that this method, that we have called DTA, provides analytical expressions for the system GFs which are as simple as previous 
analytical approximations for the polaronic regime like SPA and PTA. We show, however, that the DTA eliminates the more remarkable pathologies
of these two previous approximations in the low energy (SPA) and high energy (PTA) regimes. By comparison with other methods we have shown
that DTA additionally reproduces the correct behavior in the crossover regime, $\lambda^2/\omega_0 \lesssim \Gamma$. Only in the limit
$\lambda^2/\omega_0\Gamma \ll 1$ this approximation progressively deviates from the results provided by perturbation theory in the
electron-phonon coupling. Some exact known limits of the model like the fully empty or fully occupied dot case are also recovered 
within DTA. 

On the other hand, we have shown that DTA provides results for the current and the differential conductance in good agreement with
results from other more elaborate methods including numerically exact methods like diagrammatic Monte Carlo. In addition DTA can be
formulated in a way which allows to extract the noise properties and the more generally the FCS of the model. We have provided 
an analysis of the main features of the voltage-dependent shot noise and also of the thermal noise. For this last case we have
shown analytically that DTA fulfills the fluctuation-dissipation theorem. 
We have also demonstrated that DTA satisfies current and noise conservation
laws. This is a remarkable property in view of the simplicity of the
approximation and its non self-consistent character \cite{self-consistency}.

For future applications, the simplicity of the method could allow to address more complex situations like the non-stationary 
response of the model as studied in Ref. \cite{Ferdinand}. 
On the other hand, although the method has been derived for the more
simple single model, the same ideas could be in principle extended to
models including several dot levels coupled to multiple phonon modes
like the one discussed in Ref. \cite{brandes}. One can also envisage
improving the present approximation by including non-equilibrium
effects in the phonon distribution, as discussed for instance in
Ref. \cite{galperin}.

\begin{acknowledgments}
We would like to thank A. Zazunov and R. Avriller for very useful discussions. We are in debt to K. F Albrecht for sending us the quantum Monte Carlo 
data for the comparisons. We also thank Spanish MINECO 
for financial support under project FIS2011-26516.

\end{acknowledgments}

\appendix
\section{Expressions for the noise and the CGF in the different approximations}
\label{appendix}

Within PTA the CGF has the same expression as for a non-interacting
system \cite{paperPTA}
\begin{eqnarray}
 \ln\chi(\nu)&=&\int \frac{d\omega}{2\pi}\ln\{1+T(\omega)\left[\left(e^{i\nu}-1\right)f_L(\omega)\left(1-f_R(\omega)\right)\right.\nonumber\\
  &+&\left.\left(e^{-i\nu}-1\right)f_R(\omega)\left(1-f_L(\omega)
\right)\right]\} ,
\end{eqnarray}
with a renormalized transmission $T(\omega)$ given by
${T(\omega)=4\Gamma_L\;\Gamma_R/\left(f^{-2}(\omega)+\Gamma^2\right)}$, where
\begin{equation}
 f(\omega) \equiv \sum^{\infty}_{k=-\infty}\alpha_k\;
\left(\frac{n_0}{\omega-\tilde{\epsilon}+k\;\omega_0}+
\frac{1-n_0}{\omega-\tilde{\epsilon}-k\;\omega_0}\right) .
\end{equation}

For the other approximations (DTA and DSPA), the expressions  
are slightly more involved. We will first analyze the DTA case. The 
Keldysh GFs within  
this approximation including the counting field can be written as
\begin{eqnarray}
G_{DTA}^{+-(\nu)}(\omega)=-\sum_{k=-\infty}^{\infty}\alpha_k\frac{\tilde{\Sigma}_{L}^{+-}(\omega')e^{i\nu}+\tilde{\Sigma}_{R}^{+-}(\omega')}{\tilde{\textfrak{D}}^{(\nu)}(\omega')}\nonumber\\
 G_{DTA}^{-+(\nu)}(\omega)=-\sum_{k=-\infty}^{\infty}\alpha_{-k}\frac{\tilde{\Sigma}_{L}^{-+}(\omega')e^{-i\nu}+\tilde{\Sigma}_{R}^{-+}(\omega')}{\tilde{\textfrak{D}}^{(\nu)}(\omega')} , \nonumber\\
 \label{CGF_DTA}
\end{eqnarray}
where $\omega'=\omega+k\;\omega_0$ and
\begin{eqnarray}
 \textfrak{D}^{(\nu)}(\omega)=\tilde{\textfrak{D}}(\omega)+&\tilde{\Sigma}^{-+}_{0L}&(\omega)\tilde{\Sigma}^{+-}_{0R}(\omega)\left(e^{-i\nu}-1\right)\nonumber\\
+&\tilde{\Sigma}^{+-}_{0L}&(\omega)\tilde{\Sigma}^{-+}_{0R}(\omega)\left(e^{i\nu}-1\right) .
\end{eqnarray}

By using Eq. (\ref{generalI}) for the current between the dot and the right
electrode one obtains 
\begin{equation}
 I_{DTA}=\int\frac{d\omega}{2\pi}\frac{\tilde{\Sigma}^{+-}_{0R}(\omega)\tilde{\Sigma}^{-+}_{0L}(\omega)-\tilde{\Sigma}^{-+}_{0R}(\omega)\tilde{\Sigma}^{+-}_{0L}(\omega)}{\tilde{\textfrak{D}}(\omega)},
 \label{Ir_DTA}
\end{equation}
which is clearly anti-symmetric with respect to the interchange 
of $L$ and $R$, thus indicating current conservation within DTA.

In an analogous way, the voltage dependent noise can be straight-forwardly 
computed from Eq. (\ref{generalnoise})
\begin{eqnarray}
&&S_{DTA}=\int\frac{d\omega}{\pi}\frac{\tilde{\Sigma}^{+-}_{0R}(\omega)\tilde{\Sigma}^{-+}_{0L}(\omega)+\tilde{\Sigma}^{-+}_{0R}(\omega)\tilde{\Sigma}^{+-}_{0L}(\omega)}{\tilde{\textfrak{D}}(\omega)} \nonumber\\
&&-\left[\frac{\tilde{\Sigma}^{+-}_{0R}(\omega)\tilde{\Sigma}^{-+}_{0L}(\omega)-\tilde{\Sigma}^{-+}_{0R}(\omega)\tilde{\Sigma}^{+-}_{0L}(\omega)}{\tilde{\textfrak{D}}(\omega)}\right]^2 .
\end{eqnarray} 

In addition to current conservation
it is interesting to check whether this approximation fulfills the
fluctuation-dissipation theorem, which relates the thermal noise
at zero bias with the linear conductance by 
$S(0)=4TG(0)$. At zero voltage  $\Sigma_{0L}^{\alpha\beta}(\omega)/\Gamma_L=\Sigma_{0R}^{\alpha\beta}(\omega)/\Gamma_R$ 
($\alpha,\beta=+-$). Then, the zero voltage noise can be written as
\begin{equation}
 S_{DTA}(0)=\int\frac{d\omega}{\pi}\frac{\tilde{\Sigma}^{+-}_{0R}(\omega)\tilde{\Sigma}^{-+}_{0L}(\omega)+\tilde{\Sigma}^{-+}_{0R}(\omega)\tilde{\Sigma}^{+-}_{0L}(\omega)}{\tilde{\textfrak{D}}(\omega)}.
\label{noise-v0}
\end{equation}

On the other hand, the zero bias conductance can be computed from 
Eq. (\ref{Ir_DTA}) yielding
\begin{equation}
 G_{DTA}(0)=\frac{1}{4T}\int\frac{d\omega}{\pi}\frac{\tilde{\Sigma}^{+-}_{0R}(\omega)\tilde{\Sigma}^{-+}_{0L}(\omega)+\tilde{\Sigma}^{-+}_{0R}(\omega)\tilde{\Sigma}^{+-}_{L}(\omega)}{\tilde{\textfrak{D}}(\omega)}.
\end{equation}

Comparison with Eq. (\ref{noise-v0})  
clearly shows the fulfillment of the fluctuation-dissipation
theorem within DTA.

Similar expressions can be derived for the DSPA. First, the current flowing through the right electrode can be written as
\begin{equation}
 I_{R,DSPA}=\int\frac{d\omega}{2\pi}\frac{\Sigma^{+-}_{0L}(\omega)\tilde{\Sigma}^{-+}_{0R}(\omega)-\Sigma^{-+}_{0L}(\omega)\tilde{\Sigma}^{+-}_{0R}(\omega)}{\left(\omega-\tilde{\epsilon}\right)^2+\Gamma^2}.
 \label{Il_DSPA}
\end{equation}

It is important to notice that this expression is not symmetric with respect to $L$ and $R$ exchange when the level $\tilde{\epsilon}$ is displaced from zero. 
This absence of symmetry breaks the current conservation for bias voltages bigger than $\omega_0$.

In the same way the noise can be computed
analytically within DSPA 
\begin{eqnarray}
S_{R,DSPA}&=&\int\frac{d\omega}{\pi}\frac{\tilde{\Sigma}^{+-}_{0R}(\omega)\Sigma^{-+}_{0L}(\omega)+\tilde{\Sigma}^{-+}_{0R}(\omega)\Sigma^{+-}_{0L}(\omega)}{\Gamma^2+(w-\tilde{\epsilon})^2} \nonumber\\
&& +4\Gamma_L\Gamma_R\left(f_L(\omega)-f_R(\omega)\right) \nonumber\\ 
&& \times \frac{\tilde{\Sigma}^{+-}_{0R}(\omega)\Sigma^{-+}_{0L}(\omega)-\tilde{\Sigma}^{-+}_{0R}(\omega)\Sigma^{+-}_{0L}(\omega)}{\left[\Gamma^2+(w-\tilde{\epsilon})^2\right]^2} ,
\end{eqnarray} 
which again exhibits explicitly the breaking of left-right symmetry, i.e. noise is not conserved in general within DSPA.
It can be shown, however, that within this approximation the fluctuation-dissipation theorem is fulfilled.

\end{document}